\documentclass[aps,pre,twocolumn,groupedaddress]{revtex4-1}

\usepackage{lipsum}
\usepackage{graphicx}
\usepackage{amsmath,amssymb}
\usepackage{bm}
\usepackage{appendix}
\usepackage{dcolumn}
\usepackage{theorem}
\usepackage{wrapfig}
 \usepackage{mathtools}
\usepackage{xcolor}
 \usepackage{txfonts}
\mathtoolsset{centercolon}
\usepackage{amsmath}
\usepackage{graphicx}
\usepackage{bm}
\usepackage{float}
\usepackage{epstopdf}
\usepackage{xspace,colortbl}
\usepackage{gensymb}
\usepackage{amssymb}
\usepackage{epsfig}
\usepackage{graphicx}
\usepackage{amsmath}
\usepackage{bm}
\usepackage{color}
\makeatletter

\begin{document}

	\title [Influence of  lunisolar tides on   plants]
{Influence of  lunisolar tides on   plants.  \\
	Parametric resonance induced by periodic variations of   gravity.}

\noindent {\color{blue} AIP PUBLISHING. PHYSICS OF FLUIDS ACCEPTED MANUSCRIPT. 
This is the author's peer reviewed, accepted manuscript. However, the on line version of record will be different from this version once it has been copyedited and typeset.
PLEASE CITE THIS ARTICLE AS DOI:10.1063/5.0023717}

\author{Henri Gouin$^{1*}$ }

\affiliation{
	$^1$Aix Marseille Univ, CNRS,  
	IUSTI \, UMR 7343, Marseille, France}

\date{\today}
\email{henri.gouin@univ-amu.fr; henri.gouin@yahoo.fr}

\begin{abstract}
 	Recent experiments conducted in the International Space Station highlight the apparent periodicity of leaf oscillations and other biological phenomena associated with rhythmic variations of  lunisolar forces.  These events are similar to those occurring on Earth, but with greater effects over a shorter period of time.\\
 Among the possible disturbances, other than forced or self-existing oscillations, parametric resonances appear caused by a small periodic term;
 such is the case of fluids subjected to small periodic variations in gravitational forces in microscopic or mesoscopic plant channels filled with sap and air-vapor.        The interface instabilities verify a Mathieu's second order differential equation resulting from a    Rayleigh-Taylor stability model.  These instabilities  appear during the Moon's rotation around the Earth and during the revolution of the International Space Station. They  create impulses of pressure and sap movements   in the network of roots, stems and leaves.    
The model can explain the effects of the lunar tide on plant growth. The eccentricity of the lunar orbit around the Earth creates an important difference between the apogee and perigee of the Moon's trajectory and therefore the tidal effects can depend on the distance between the Moon and the Earth.\\

\noindent PACS Numbers: 87.10.Ed, 87.18Yt, 87.16.dj, 96.20.Jz, 96.25De

\end{abstract}
  
\maketitle

 \section{\label{Introduction} Introduction and experimental evidences}
 The mechanisms involved in the treatment of lunisolar  field  on plants are not fully understood or even accepted   \cite{Moran}.  The existence of a  diurnal rhythm of leaf movement  has been observed since ancient times and has been considered for a long time  as a circadian cycle    \cite{Bunning3,Tavener,Engelmann}.

 About 30 per cent of   tidal forces is due to   the Sun and 60 per cent is due to   the Moon. The variation in gravitational forces creates these lunisolar tides on  Earth.  
 In  literature, two competing models currently exist on how the motion of water in and out of cells may create leaf motions:  
the first is based on Dorda's hypothesis in which gravity, mass and time are treated in quantified form \cite{Dorda}; in the second, Jurin's law of capillarity is used    \cite{Gonen}. 
 However, it seems these  models do not create  strong enough   impulses    due to gravitational variations. \\
 Barlow examined bean-leaf movement data from the 1920s to 2015 \cite{Barlow3}. Even when  light and dark periods  overlap the time intervals, the duration of a  cycle of motions is 24.8 hours  \cite{Klein,Fisahn1}.
 Due to the   relative orbital motions of the Earth and the Moon around the Sun, the Earth's gravitational field is continuously changing  \cite{Konopliv1,Konopliv2}. These modulations induce  the diurnal rise and fall of ocean tides,  small elastic deformations of the Earth's crust  \cite{Gallep1,Gallep2,Harmer} and
 for  expanding shoots and roots  \cite{James}.   The influence is noticeable on plant growth  and stem elongations, as well as variation in  root diameters  
 \cite{Beeson,Sorbetti-Guerri,Zajaczhowska,Xu,Crossley}.
 The Moon  distance to the Earth varies   along its orbit and  is  perceptible  on living organisms  \cite{Palmer,Miles,Gluckman}.

Modern tools allow  observations in the   International Space Station (ISS). The cycles of leaf motions are     aligned with the revolution time on its orbit   \cite{Brinckman,Bobst,Jacob}. 
 The rhythm of   lunisolar tidal forces in the ISS is very different from the rhythm   on Earth:    the ISS orbits the Earth approximatively  every 90 minutes. Rhythms inside the ISS  with two high and two low lunisolar tides per orbit create oscillations of \textit{Arabidopsis thaliana} leaves   with periods of 45  minutes  \cite{Solheim,Fisahn2}.

Our study consists of  \textit{toy models} made of cylindrical microscopic or mesoscopic tubes with diameters a few tenths to a few tens of microns corresponding to the diameters of the xylem channels in roots and stems. Channels are filled with  sap  and gas. The first  \textit{toy  model}  consists of horizontal  microscopic or mesoscopic channels. Thanks to capillary energy, the sap wets  channel edges.  The fluid consisting of gas  is located in   central part of the cylinder, creating an interface.   The second  \textit{toy  model}  consists  of vertical  microscopic or mesoscopic tubes, simulating  on Earth xylem-channels of plant stems. According to Jurin's law the lower part of the tube is filled with sap and the upper part with air-vapor.  
\\
A Rayleigh--Taylor model at constant gravity   can study the stability   of liquid-sap/air-vapor interfaces in the  channels  at equilibrium \cite{Chandrasekhar,Charru}. \\  To better understand the purpose of the article, let us give an   example of instability:  
a vertical pendulum of period $T_0$ can be set in motion by rhythmically increasing and decreasing its length with a period commensurable with $T_0$.  
The length $\ell$ of the pendulum varies according to
$
\ell =\ell_0(1+f(t))$, ($\ell_0$ constant), 
where $|f(t)|\ll 1 $  is a periodic  function of period $T_0/q$  when $q$ is a small integer.
This amazing phenomenon is called \textit{parametric resonance} \cite{Arnold2}. \\
 As in the example of the pendulum, but due to small periodic variations in gravity, the harmonic-oscillator equations obtained by the Rayleigh-Taylor model taken for \textit{toy models} are modified and create parametric resonances. These results come from Mathieu's equations whose points of instability are analyzed in relation to the revolutions of the Moon and the ISS around the Earth.\\   In this paper, we theoretically prove parametric resonances associated with gravity variations induce sap and pressure impulses corresponding to plant observations made on Earth and in the ISS.

 	\section{Stability of fluid-interfaces in     microscopic or mesoscopic tubes submitted to constant gravity} 

 \subsubsection{Equilibrium preliminaries}
 
 To understand   gravity  effects on  stability of  microscopic or mesoscopic channels, we first consider the simple  \textit{toy-model}  of a cylindric  tube filled with liquid and gas separated by a liquid/gas interface. The liquid is crude or elaborated sap; the gas is air and   vapor. The tube is assumed to be horizontal on Earth and in any position inside  the ISS. We first consider the case of  tubes on   Earth (see Fig. \ref{Fig.1}).\\
 Because  the fluid energy decreases when  crude or elaborated saps wet  xylem walls, the  sap  wets  channel walls   \cite{Mattia,Gouin0}. As in case of stability problems for the Rayleigh-Taylor model, the fluids are supposed to be incompressible \cite{Chandrasekhar,Charru}.   

 The reference state is    equilibrium, where a quantity $\beta$ is then referred  as $\overline \beta$. We have:
 \begin{equation*}
 \overline {P_a} = \overline {P_{a0}}-\rho_a\,g\, y  \quad{\rm and}\quad
 \overline {P_b} = \overline {P_{b0}}-\rho_b\,g\, y  
 \end{equation*} 
 where $\overline {P_a}$ and $\overline {P_b}$ are the fluid pressures in  $(a)$ and $(b)$ domains at   $y$ level, $\rho_a$ and $\rho_b$ are the fluid densities, $\overline {P_{a0}}$ and $\overline {P_{b0}}$ are the pressures at   level $y=0$,   and $g$ is the gravity acceleration assumed  constant.     At level $y=0$, we have Laplace's relationship\
$
 \overline {P_{a0}}-\overline {P_{b0}}=  {\gamma}/{r_{a0}}
$\
 where $r_{a0}$ is the associated radius at level $y=0$,  ${\gamma}$ is the surface energy of the liquid-sap
/air-vapor interface; $ \overline {P_{b0}}= \overline {P_{0}}$ denotes the reference pressure.  In  $(a)$ and $(b)$ domains,  equilibrium equations are written:
 \begin{equation}
 \overline {P_a} - \overline {P_{0}}=\displaystyle\frac{\gamma}{r_{a0}}-\rho_a\,g\, y \quad{\rm and}\quad
 \overline {P_b} -\overline {P_{0}}=-\rho_b\,g\, y   
 \label{RP}
 \end{equation} 
 \subsubsection{Linear   perturbations at  interface}
\begin{figure}
	\begin{center}
		\includegraphics[width=7.5cm]{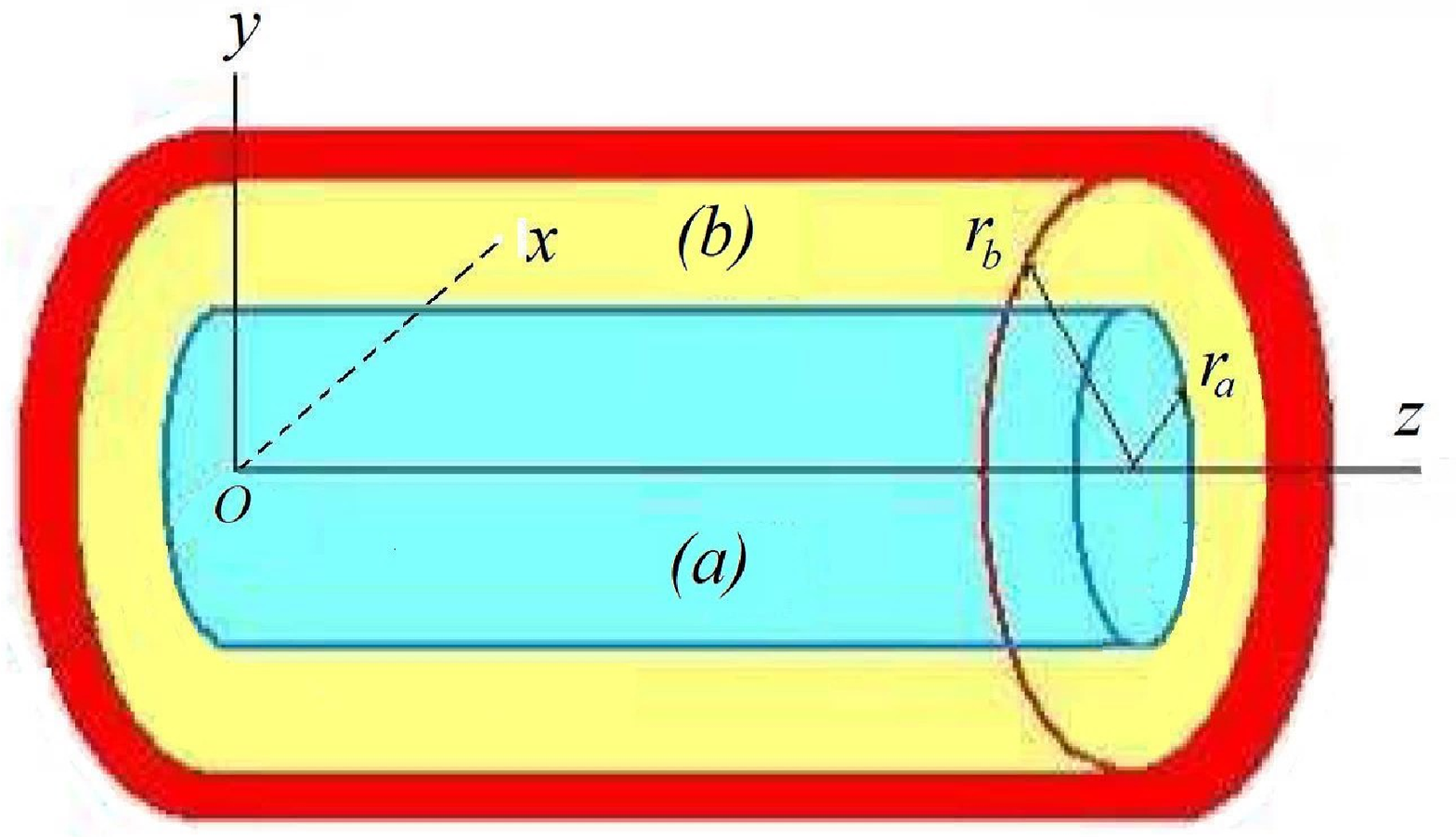}
	\end{center}
	\caption{Cylindrical microscopic or mesoscopic tube:  domain  (a) of radius  $r_a$    is occupied by air-vapor;  domain (b)  of external radius  $r_b$ and internal radius $r_a$ is occupied by   liquid-sap; the liquid-sap/air-vapor  interface is the cylinder separating (a) and (b). Axis $x$ is horizontal, axis $y$ gives the upward direction of gravity and axis $z$ is the cylinder axis.}\label{Fig.1}
\end{figure}

 Out from   equilibrium, we denote by $ {\boldsymbol{u}_a} $ and $ {\boldsymbol{u}_b} $ the fluid velocities in  $(a)$ and $(b)$ domains, and we write the pressure perturbations $p_j$:
 \begin{equation*}
 P_j = \overline {P_{j}}+p_j  \qquad  j\in\{a, b\} \end{equation*}
 For   Rayleigh--Taylor's model, the stability problem can be solved by   perturbations of  velocities.  As demonstrated in \cite{Chandrasekhar,Charru}, it is enough to consider  irrotational   velocities:
 \begin{equation}
 {\boldsymbol{u}_j} = {\rm grad}\, \phi_j  \qquad   j\in\{a, b\}\label{grad}
 \end{equation}
 where  functions $ \phi_j$ are expressed in cylindrical coordinate system $(r,\theta,z)$. The conditions of incompressibility  of fluids are written:
 \begin{equation}
 \frac{1}{r}\frac{\partial \phi_j}{\partial r}+\frac{\partial^2 \phi_j}{\partial r^2}+\frac{1}{r^2}\frac{\partial^2 \phi_j}{\partial \theta^2}+\frac{\partial^2 \phi_j}{\partial z^2} =0\qquad   j\in\{a, b\}\label{incompresible}
 \end{equation}
To consider mass   transfert, we can refer to \cite{ Seadawy}.
 We also assume that the viscosities are negligible \cite{Chandrasekhar,Charru}. 
 Effect of viscosity   are considered in \cite{Hu}. The  conservation of  fluid momenta  in   $(a)$ and $(b)$ domains are given by Bernoulli's equations  which are  first integrals of Euler's equations   \cite{Landau}:
 \begin{equation}
 \left\{\begin{array}{l}
 \displaystyle\rho_a \frac{\partial\phi_a}{\partial t}+\frac{1}{2}({\rm grad} \phi_a)^2+\overline{P_a}+p_a+\rho_a\, g\, y=\overline {P_{0}}+\displaystyle\frac{\gamma}{r_{a0}}\\
 \\
 \displaystyle\rho_b \frac{\partial\phi_b}{\partial t }+\frac{1}{2}({\rm grad} \phi_b)^2+\overline{P_b}+p_b+\rho_b\, g\, y=\overline {P_{0}} \\
 \end{array}\right.\label{momentum}
 \end{equation}
The equations   are satisfied at equilibrium,  terms $\overline {P_{0}}$ and $\displaystyle\overline {P_{0}}+\displaystyle {\gamma}/{r_{a0}}$\,\
 appear in   \eqref{momentum}.  Difference  between  \eqref{RP} and  
 \eqref{momentum} yields:
 \begin{equation}
 \left\{\begin{array}{l}
 \displaystyle\rho_a \frac{\partial\phi_a}{\partial t}+\frac{1}{2}( {\rm grad} \phi_a)^2+ p_a = 0\
 \\
 \\
 \displaystyle\rho_b \frac{\partial\phi_b}{\partial t }\,+\frac{1}{2}({\rm grad} \phi_b)^2\,+ p_b =0 \\
 \end{array}\right.\label{momentumbis1}
 \end{equation}
  The magnitude order of   tube  diameters  is a few tenths to a few tens of microns.  Due to     channel   sizes  much less than the capillary length, the shapes of  interfaces are independent of  the gravity \cite{degennes}.  Therefore, the shapes of liquid-sap/air-vapor interfaces are cylindrical around  $z$--axis    (see Fig. \ref{Fig.1}).   \\

 The  perturbed liquid-sap/air-vapor interface $H$ of equation $H(r,\theta,z,t)=0$ can be written:  
 \begin{equation*}
 H(r,\theta,z,t)=r-r_a-\eta 
 \end{equation*}
 where $\eta=\eta(\theta, z, t)$ denotes the dynamical displacement of the  liquid-sap/air-vapor interface. From calculations obtained in Appendix A.1, we obtain  
 Laplace's relationship expanded to  first order with respect to $\eta$ :
 \begin{equation}
 P_a(\eta) -P_b(\eta) =\gamma\left\{ \frac{1}  {r_a}     -\frac{1}{r_a^2}     \left(\eta+ \eta^{\prime\prime}_{\theta^2}\right)-        \eta^{\prime\prime}_{z^2}\right\}   +\mathcal{O}(\eta^2)\label{Laplace}
 \end{equation}
  where $^{\prime\prime}$ denotes the second order partial derivative with respect  to   $\theta$ or $z$,
 \subsubsection{Study of   perturbations}
 The linearization of  \eqref{incompresible} and \eqref{momentumbis1} at interface $r=r_a$  leads to a system with constant coefficients in $(z, t)$. The dependence of solutions in $z$ and $t$   is therefore exponential,
 and disturbances can be researched   in    normal modes of form \cite{Ruggeri}:
 \begin{equation}
 \left\{\begin{array}{l}
 {\phi_j}= \hat{\phi_j}(r, \theta)   \, e^{i(kz-\omega t)}
 \\ \\
 {p_j}= \hat{p_j}(r, \theta)   \, e^{i(kz-\omega t)}\quad\qquad j\in\{a, b\}\label{Pert}
 \\ \\
 {\eta}= \hat{\eta}(\theta)   \, e^{i(kz-\omega t)}\quad\,\,\ \quad\qquad {\rm at}\quad r=r_a\\
 \end{array}\right.
 \end{equation}
 where $k  \in \mathbb{R}^{+\star}$ is the wave number,  $\omega  \in \mathbb{R}^{+\star}$ is the wave frequency. We denote  by  $\lambda = \displaystyle {2\pi}/{k}$   the wavelength. 
 \\
 At  interface, we look for   perturbations $\eta$\,\ corresponding to   wave frequency $\omega$ and we deduce:
 \begin{equation*}
 {\eta}(z,t)= \hat{\eta}\left(\theta\right)  \, e^{i(kz-\omega t)}\quad\Longrightarrow\quad \ddot{\eta}(z,t)= -\omega^2{\eta}(z,t)
 \end{equation*}
 where\,\ $\ddot{}= d^2/dt^2$\,\ denotes the second order particular derivative.
 Then,  perturbation ${\eta}$   verifies the differential equation:
 \begin{equation}
 \ddot{\eta}+\omega^2\,\eta =0 \label{etapert}
 \end{equation}

 \noindent  Likewise, we denote $[p]$ the discontinuity  at   interface liquid-sap/air-vapor of pressure $p$:
 $$
 [p(z, t)]=  \left(\,\hat{p_a}(r_a, \theta)-\hat{p_b}(r_a, \theta) \right)  \, e^{i(kz-\omega t)} 
 $$ 
 Then,
 \begin{equation}
 [\ddot p]+\omega^2\,  [p]=0 
 \label{ppert}
 \end{equation} 
 \\
To be complete, in Appendix A.2, we prove that the velocities $\boldsymbol{u}_j$\,\ in  relation \eqref{grad} and perturbations $\eta$\,\ are compatible with motion equations \eqref{momentumbis1} and boundary conditions.

 \subsubsection{Numerical applications and their consequences}

 The fluid in $(a)$ domain is air and vapor; the fluid in $(b)$ domain is crude or elaborated sap. The temperature is generally about $20\degree$ Celsius.   
 In \textit{\textbf{C.G.S.}} units, we have  \cite{Weast}:\\
 
 \noindent For the crude sap (which is mainly water with some salts):
 $$\rho_a\simeq 0, \quad \rho_b \simeq 1,\quad  \gamma\simeq 70$$
 For the elaborated sap (for example maple syrup):
 $$\rho_a\simeq 0, \quad  \rho_b \simeq 1,34,\quad  \gamma\simeq 70$$
 For example, the  sizes in the micro-channel are assumed to be: 
 $$r_a= 2\times 10^{-3} {\rm cm} = 20\,\mu {\rm m},\quad  r_b= 4\times 10^{-3} {\rm cm} = 40\,\mu {\rm m}$$ 
 For mesoscopic diameters, the obtained results will be even  more accurate. 
The  equation verified by values $k$ and $\omega$ of perturbations \eqref{Pert} is the result of Appendix A.2; we obtain: 
\begin{equation}
\begin{array}{l}
\displaystyle \omega^2\,\left(r_a\,\rho_a+\frac{r_b^{2n}r_a^{-n}+r_a^n}{r_b^{2n}r_a^{-n-1}-r_a^{n-1}}\,\rho_b\right)=\\  \displaystyle \qquad\qquad\qquad \gamma\,n\,\left(k^2+ \frac{n^2-1}{r_a^2}\right)-n\,\left(\rho_b-\rho_a\right)\, g\, {\rm sin }  \theta 
\label{mother equation}
\end{array}
\end{equation}
which  is an extension of the  Rayleigh--Taylor equation  for planar interfaces \cite{Charru}.\\  

 $\bullet\quad $ if $n \geq 2, \ n\in \mathbb N$, then 
 $   {(n^2-1)}/{r_a^2}> 7.5\,\times 10^5 $.
 For any value of $g \  ({\rm where} \ g < 10^3  \  {\rm cm.s}^{-2})$, 
 $$
 (n^2-1)\,\frac{\gamma}{r_a^2}> 5 \times 10^7 \gg  \left(\rho_b-\rho_a\right)g\, \mid {\rm sin}\theta\mid 
 $$
 Consequently, for $n \geq 2$, in     \eqref{mother equation}, the gravity has no influence on the perturbations along the interface. From \eqref{mother equation} we obtain: 
 \begin{equation*}
 \gamma\,n\,k^2=\omega^2\left(\frac{r_b^{2n}r_a^{-n}+r_a^n}{r_b^{2n}r_a^{-n-1}-r_a^{n-1}}\,\rho_b\right)-\gamma\,n\,\left(  \frac{n^2-1}{r_a^2}\right) 
 \label{mother equation2}
 \end{equation*}
 The  values of $\omega$ such that 
 \begin{equation*}
 \omega^2\left(\frac{r_b^{2n}r_a^{-n}+r_a^n}{r_b^{2n}r_a^{-n-1}-r_a^{n-1}}\,\rho_b\right)-\gamma\,n\,\left(  \frac{n^2-1}{r_a^2}\right) >0
 \end{equation*} 
 are only possible. It can be immediately seen  that $\omega$    corresponds to a period $T\ll 1$  s. As we will see in Section III, this period   is not in our study range.
 \\
 
 $\bullet\quad$ if $n =1$, the  equation \eqref{mother equation}  yields:
 \begin{equation}
 \omega^2\, r_a \frac{r_b^2+r_a^2}{r_b^2-r_a^2}\,\rho_b=\gamma\, k^2 - \,\rho_b \, g\, {\rm sin } \, \theta \label{frequency}
 \end{equation}
 It is necessary that 
 $\gamma\, k^2$ and  $\rho_b\,g\,  {\rm sin }\,  \theta$ are of the same order of magnitude. The most significant case corresponds to  the micro-tube top where the effect of gravity is maximum: 
 $$
 {\rm sin }\,  \theta=1\quad\Longleftrightarrow\quad \theta=\frac{\pi}{2}
 $$
 In this case, the terms $\gamma\, k^2$ and $\textbf\,\rho_b \, g$ are equal when the value   $k=2\pi/\lambda$ is minimum equal to $k_m$ corresponding to $\lambda$ maximum and equal to $\lambda_M$, \textit{i.e.}:
 \begin{equation*}
 k_m =\sqrt{ \frac{\rho_b
 		\, g}{\gamma}} \quad\Longleftrightarrow\quad \lambda_M =2\pi\sqrt{ \frac{ \gamma}{\rho_b
 		\, g} }
 \end{equation*}
 For a stable equilibrium position when $g$ is constant,    $\omega^2$  must be positive. Consequently, due to the fact $r_b>r_a$ and $\omega^2>0$,
 \begin{equation*}
 \gamma\, k^2 -\rho_b
 \, g>0
 \end{equation*}
  For a liquid--gas plug length of less than $\lambda_M$, the fluids are stable in the Rayleigh--Taylor model. The oscillations have an angular frequency $\omega$ given by \eqref{frequency}. The period of oscillation is $T =2\pi/\omega$. We obtain from equation \eqref{mother equation}:
 \begin{equation*}
 k^2= {\frac{\rho_b g}{\gamma}\left(1+\frac{\omega^2\,\gamma\,r_a\left(r_b^2+r_a^2\right)}{g\left(r_b^2-r_a^2\right)}\right)} \equiv{\frac{\rho_b g}{\gamma}\left(1+\frac{4\pi^2\,\gamma\,r_a\left(r_b^2+r_a^2\right)}{g\,T^2\left(r_b^2-r_a^2\right)}\right)}\label{mother2}
 \end{equation*}
 Let us consider the case of an horizontal micro-channel on  Earth.
 For $T> 10^3$ s, $r_a<0.01$ cm, $r_b=1.1\, r_a$, then
 \begin{equation*}
 \frac{\displaystyle 4\pi^2\gamma\,r_a}{\displaystyle g\,T^2}\frac{\displaystyle r_b^2+r_a^2 }{\displaystyle r_b^2-r_a^2} < 3\times 10^{-7}  
 \end{equation*}
 is negligible with respect to 1.
 On Earth, $g=981\, {\rm  cm. s}^{-2}$, we obtain $\lambda_M\simeq1.68$ cm. This is the maximum size of  fluid  plug lengths.

 We may extend the previous results for micro-channels inside the  ISS.    
 In    ISS revolutions, the  apparent gravity  is less than $150\,\mu {\rm Gal}$  \cite{Barlow3}.   It is   easy to verify that $\lambda_M$ is over  hundred  meters and the interface is always stable in Rayleigh--Taylor's model. The perturbation of   interfaces are always in form \eqref{Pert};   calculations are similar, and  \eqref{etapert} and \eqref{ppert} are always satisfied.\\

 \subsection{Stability of  vertical micro-tubes}
\begin{figure}
 	\begin{center}
 		\includegraphics[width=3.5cm]{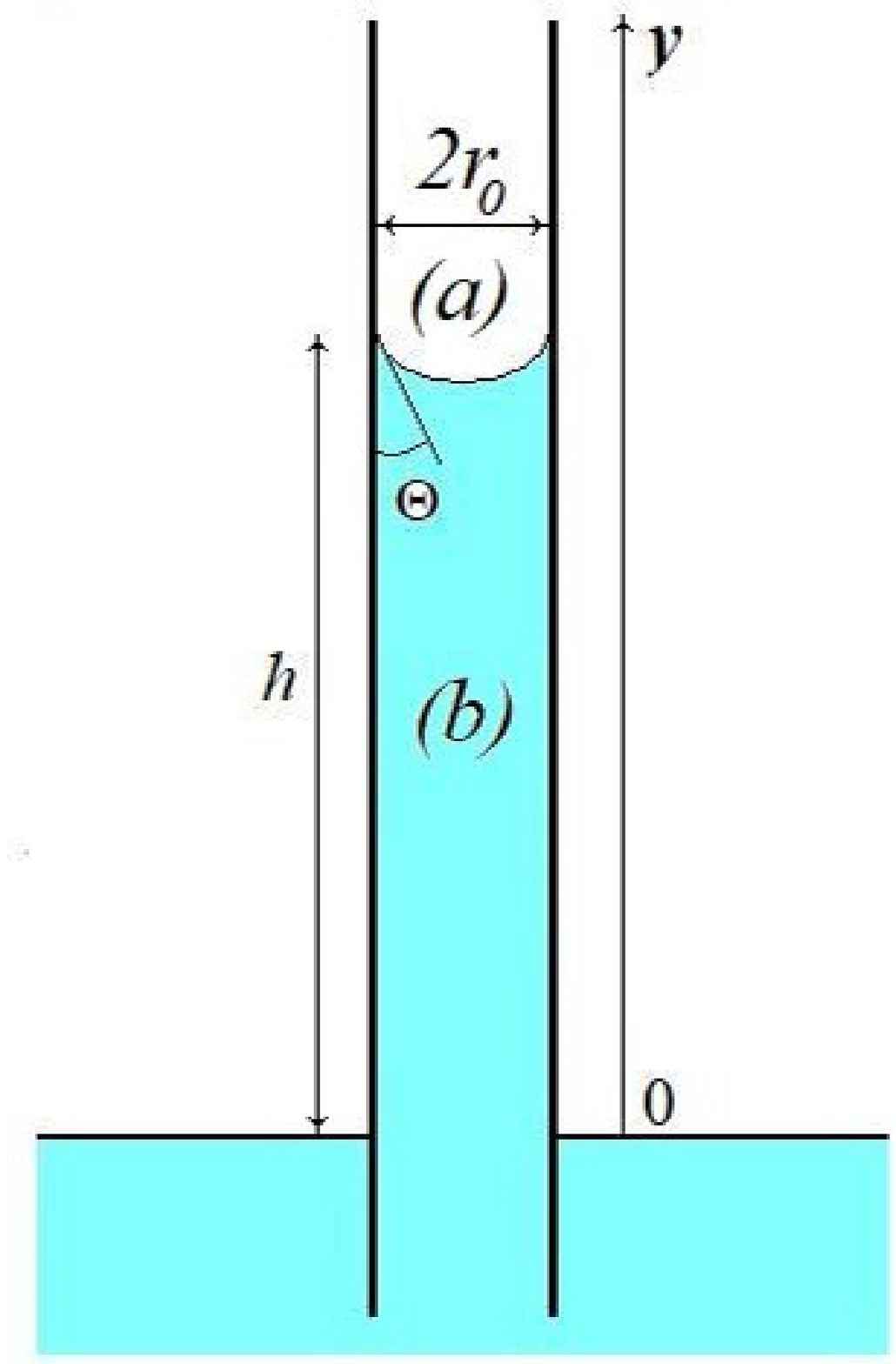}
 	\end{center}
 	\caption{A vertical cylindrical micro-tube of  radius $r_0$  is filled until level $h$ by the crude sap. Sap constant density  is $\rho_b$; $\Theta$ is the Young contact angle of the meniscus    with the micro-tube's wall.} \label{Fig.2}
 \end{figure}

 This case   corresponds  to stem  xylem-channels on Earth. 
 We  consider a  second   \textit{toy-model} constituted by a vertical  cylindrical   tube (with diameters a few tenths to a few tens of microns),  filled with liquid-sap and air-vapor.  The   tube   is connected   with a liquid-sap reservoir   (see Fig. \ref{Fig.2}).        
 The liquid-sap rises at   level $ h$ and creates a  liquid-sap/air-vapor interface  (\textit{meniscus}).  
 Jurin's law of capillarity writes:
 \begin{equation}
 h = \frac{2\,\gamma\,  {\rm cos}\,\Theta}{r_0\,\rho_b\,g} \label{Jurin}
 \end{equation}
 where  $\rho_b$ is the  liquid density and  $ \gamma$ is the sap surface tension. 
 We note that if $g$ varies infinitesimally, this has no consequence on  $h$-height of the meniscus. In fact \eqref{Jurin} is only the equilibrium equation for fluids in   tubes. Dynamic perturbations of the  meniscus must be studied  from  equations of motions. 
 We again assume  that   fluids are incompressible and with negligible viscosities; we consider  meniscus perturbations  $\eta$ in   vertical direction. The calculus are simpler than in Section 2.1;    we  refer to Chapter 2 in \cite{Charru}.\\
 The linearization of   perturbation equations yields a similar form as \eqref{incompresible} and \eqref{Laplace} for  meniscus displacements; it comes a system with  coefficients 
 independent of $r$ and $t$, where $r$ is the radial coordinate. The dependence of solutions in $r$ and $t$   is therefore exponential,
 and disturbances can be researched   in   form of normal modes.  We  obtain a representation in a  form similar to  Eq. (2.34) in \cite{Charru},   but with $r$ in place of $x$:  
 \begin{equation*}
 \left\{\begin{array}{l}
 {\phi_j}= \hat{\phi_j}(y),  \, e^{i(kr-\omega t)}
 \\ \\
 {p_j}= \hat{p_j}(y)   \, e^{i(kr-\omega t)}\qquad j\in\{a, b\}, \  k \in \mathbb{R}^{+\star}\ {\rm and} \,\ \omega\in \mathbb{R}^{+\star}\label{TubePert}
 \\ \\
 {\eta}= \hat{\eta}(y)   \, e^{i(kr-\omega t)}\\
 \end{array}\right.
 \end{equation*}
 where domain $(a)$ corresponds to  air-vapor and domain $(b)$ to   liquid-sap. 
   Perturbations ${\eta}$  at the interface again verify the differential equation:
 \begin{equation*}
 \ddot{\eta}+\omega^2\,\eta =0 \label{etapert1}
 \end{equation*}
 Discontinuity of pressure  $[p]$ through the meniscus verifies:
 $
 [p(z, t)]=  \left(\,\hat{p_a}-\hat{p_b} \right)\, e^{i(kz-\omega t)} 
 $ 
 and consequently:
 \begin{equation*}
 [\ddot p]+\omega^2\,  [p]=0 
 \label{ppert1}
 \end{equation*}
 In the case of   revolution  of the Moon around the Earth, $g=981\, {\rm cm. s}^{-2}$ and we always obtain the capillary length  $\lambda_M\simeq1.68$ cm \cite{Charru}.   Due to     radius sizes of   micro-channels,  when $g$ is assumed constant,   liquid-sap plugs always exist   and           meniscus is stable for the   Rayleigh--Taylor model. 
 	\section{Variations of  gravity due to\\ the  Moon and   Sun}
Relatively to fixed directions with respect  to  stars,  the period $T$ of a satellite around the Earth and its corresponding angular frequency $\displaystyle \omega$ can be written \cite{Brouwer}:
\begin{equation}
T= 2\pi \sqrt{\frac{a^3}{R^2 g}}\qquad {\rm and}\qquad \omega = \sqrt{\frac{R^2 g}{ a^3}}
\label{key0}
\end{equation} 
where $g$ is the associated averaged acceleration of gravity on Earth, $R$   the Earth radius and $a$   the apogee lenght of the satellite orbit. 
 \subsection{Variations of   gravity on the Earth}
 On Earth, when the Moon is at its zenith, the gravity   decreases by $110\, \mu$Gal (1\,$\mu${\rm Gal} $=10^{-6}$ cm.s$^{-2}$); when the Moon is at the horizon, it decreases by 170 $\mu$Gal. The action of the Sun, which is 2.17 times weaker than the Moon action, causes a maximum variation of $80 \,\mu$Gal under similar conditions. In the most favorable case  the gravity undergoes a variation of $250 \,\mu$Gal, corresponding to a variation of about  1/4 000 000 th of its intensity  \cite{Gougenheim,Longman}.  The actual estimation of  gravimetric--tide is provided by  Kingel\'e's
 \textit{Etide program}  \cite{Volkov}. However, due to  the circulation of the Moon   around the Earth over one month,   the  distance from   Moon to   Earth  varies. The spacing between the two bodies does not significantly  vary during   one day; the  value is given for a position along the lunar orbit (see Fig. \ref{Fig.3}).
 \\
 The Moon moves  around the Earth axis $O\boldsymbol{k}_0$ relatively to fixed directions with respect  to stars during $T_L = 27.3$ days. The associated angular frequency is $\displaystyle \omega_L  =2\pi\times 4.24\times 10^{-7} $  {\rm rad.s$^{-1}$.\\
 Relatively to  fixed directions with respect to   stars, the Earth rotates around axis $O\boldsymbol{k}_0$ during $T_E=$ 24 hours. The  associated angular frequency is $\omega_{_E}  =2\pi\times1.16\times 10^{-5} $  {\rm rad.s$^{-1}$. \\
 Relatively to Earth-bound axes of third axis $O\boldsymbol{k}_0$, the apparent Moon motion  has  a period of   $\displaystyle T_{L_E} =24.8$ hours  \cite{Barlow3}. The associated angular  frequency  is 
 $\omega_{L_E}=2\pi\times 1.12\times 10^{-5}$  {\rm rad.s$^{-1}$.         \\
 The   gravity on  Earth   is not constant and its  perturbation  has a period  which is  associated with  the apparent motion of the Moon around the Earth.   A simple expression of the gravity variation on Earth can be written in periodic form as:
 \begin{equation*}
 g(t) =g_{_E}\left(1+\varepsilon_{_E} {\rm cos} \left(\omega_{L_E}\,(t-t_0)\right)\right), 
\quad g_{_E}=\frac{1}{T_{L_E}}\int_{0}^{T_{L_E}}  g(t) dt\label{garvity1} 
 \end{equation*}
 where  $\varepsilon_{_E}=1.25\times 10^{-7} =$ 1/8 000 000  with $g_{_E}=9.81 \times 10^8 \mu$Gal.
 Because $g$ is associated with the Moon orbit  in   axes centered at the Earth and  fixed  directions with respect to   stars, we obtain from \eqref{key0}:
 \begin{equation*}
 \omega_L^2 = \frac{R^2 g(t)}{ a^3}=   \frac{R^2   g_{_E}\left(1+\varepsilon_{_E} {\rm cos} \left(\omega_{L_E}\,(t-t_0)\right)\right)}{ a^3 }
 \end{equation*}
 where $a$ is the Moon apogee relative to the Earth.
 \begin{equation*}
 \omega_L^2 = \omega_{L_e}^2  \left(1+\varepsilon_{_E} {\rm cos} \left(\omega_{L_E}\,(t-t_0)\right)\right)\quad{\rm with}\quad\omega_{L_e}^2 = \frac{R^2g_{_E}}{a^3}
 \end{equation*}
 The  revolutions being associated with the rotation around the axis $O\boldsymbol{k}_0$ of the plane containing $O\boldsymbol{k}_0$  and the Moon center, then\,\   $\omega_{L_E}+\omega_L=\omega_{_E}$ and  due to  $\varepsilon_{_E}\ll 1$:
 \begin{equation*}
 \omega_{L_E}=\omega_{_E}-\omega_{L}\equiv\omega_{_E}-\omega_{L_e}  \left(1+\frac{\varepsilon_{_E}}{2} {\rm cos} \left(\omega_{L_E}\,(t-t_0)\right)\right)
 \end{equation*}
 We denote $\omega_{L_{E_0}}=\omega_{_E}-\omega_{L_e}$. Due to the fact that $\omega_{L_{E_0}}\simeq\omega_{L_E}$, we obtain:
 \begin{equation*}
 \omega_{L_E}^2=\omega_{L_{E_0}}^2  \left(1+\varepsilon_{_{E_0}} {\rm cos} \left(\omega_{L_{E_0}}\,(t-t_0)\right)\right)
 \end{equation*}
 where $\displaystyle \varepsilon_{_{E_0}} =\left(\omega_{L_e}/{\omega_{L_{E_0}}}\right) \varepsilon_{_E} =0.47\times 10^{-8}$.

 \subsection{Variations of   gravity at the ISS altitude}

 Relatively to Earth-bound axes of third axis $O\boldsymbol{k}_0$, the ISS revolution has  a period   $T_{I_E} = 93\ {\rm minutes}$. The ISS   moves  around axis $O\boldsymbol{k}_0$ with the same orientation than the Earth (trigonometric rotation), but $15.8$ times faster.   The  associated angular frequency is 
 $
 \omega_{I_E}=2\pi\times 1.79\times 10^{-4} $  {\rm rad.s$^{-1}$. 
 	
 \begin{figure}
 	\begin{center}
 		\includegraphics[width=8.5
 		cm]{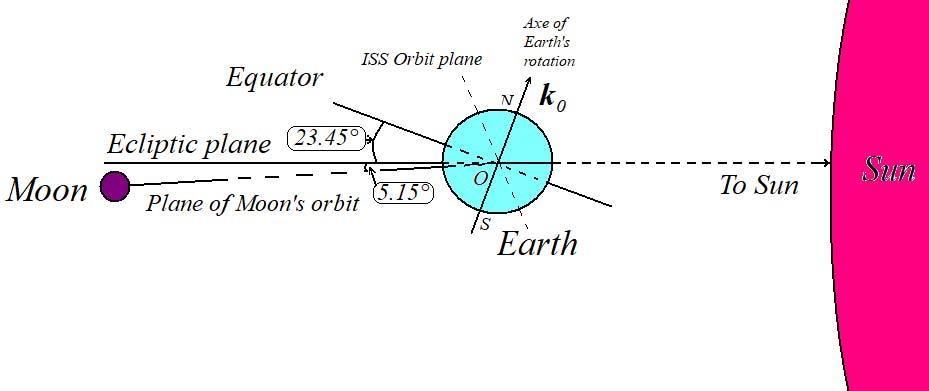}
 	\end{center}
 	\caption{The Earth's rotation around axis $\boldsymbol{k}_0$. The ISS orbit, the Moon and Sun positions are indicated on the  figure.} \label{Fig.3}
 \end{figure} 
 Relative to fixed directions   with respect  to   stars, the Earth rotates around  axis 
 $O\boldsymbol{k}_0$ in $T_E=$ 24 hours. The revolutions being associated with the rotation around the axis $O\boldsymbol{k}_0$ of the plane containing $O\boldsymbol{k}_0$  and the ISS,    the ISS has a sidereal angular frequency (frequency relative to axes of fixed direction with respect to  stars)  
 $
 \omega_I = \omega_{I_E}+\omega_E= 2\pi\times 1.91\times 10^{-4}$ {\rm rad.s$^{-1}$.
  
 Variations of the lunisolar tidal acceleration   in the ISS are closely sinusoidal \cite{Fisahn1,Fisahn2}}. The gravitational oscillations expressed by the \textit{Etide program proposed by Klingelé, can be first found in \cite{Volkov}}. The Etide program
  shows that the period of gravitational variations is about 45 minutes with two maxima of about $150\ \mu$Gal and two minima of about $-70 \  \mu$Gal.

 However, due to  the circulation of the Moon   around the Earth relative to axes of fixed directions with respect to   stars  during one month,    the  distance from   Moon to   Earth  varies. Because the spacing between the two bodies does not significantly  vary during   93 minutes, the  value is given for a position when the ISS is at an altitude of 410 km. The perturbation term   has a period which is   the half of the   revolution period around the Earth.   We can write a simple  expression of   gravity at the ISS level  (it is not the micro-gravity inside the ISS but the gravity due to the Earth-Moon-Sun): 
 \begin{equation*}
 g(t) =g_I\left(1+\varepsilon_{I} {\rm cos} \left(2\,\omega_{I_E}\,(t-t_1)\right)\right),
\  \ \, g_I=\frac{1}{T_{I_E}}\int_{0}^{T_{I_E}} g(t)\, dt\label{gravity2}
 \end{equation*} 
 $g(t)$ is a periodic function of $t$ of period $T_{I_E}/2$  
 where         $g_I=8.83 \times 10^8 \mu$Gal and
 $\varepsilon_{I}= 1.25\times 10^{-7}$. 
 \\
 The gravity variation effect  on $\omega_I$ is associated with the ISS orbit   around the Earth relatively to axes of fixed directions with respect to   stars (see Fig. \ref{Fig.3}):
 \begin{equation*}
 \omega_I^2 = \frac{g(t)}{a}=   \frac{g_I\left(1+\varepsilon_{I}\, {\rm cos} \left(2\,\omega_{I_E} (t-t_1)\right)\right)} {a }
 \end{equation*}
 where $a$ is the apogee length of the ISS with respect to the Earth and here $g(t)$ is  the value of  gravity at ISS level. 
 \begin{equation*}
 \omega_I^2 = \omega_{I_e}^2  \left(1+\varepsilon_I\, {\rm cos} \left(2\,\omega_{I_E} (t-t_1)\right)\right)\quad{\rm with}\quad\omega_{I_e}^2 = \frac{ g_I}{a}
 \end{equation*}
 Due to the fact that $\varepsilon_I\ll 1$:
 \begin{equation*}
 \omega_{I_E}=\omega_{I_e}  \left(1+\frac{\varepsilon_I}{2} {\rm cos} \left(2\,\omega_{I_E}\,(t-t_1)\right)\right)-\omega_E
 \end{equation*}
 We denote $\omega_{I_{E_0}}=\omega_{I_e}-\omega_{_E}$. From $\omega_{I_{E_0}}\simeq\omega_{I_E}$, we obtain:
 \begin{equation*}
 \omega_{I_E}^2=\omega_{I_{E_0}}^2  \left(1+\varepsilon_{I_0} {\rm cos} \left(2\,\omega_{I_{E_0}} (t-t_1)\right)\right) 
 \end{equation*}
where $ \varepsilon_{I_0} =\left({\omega_{I_e}}/{\omega_{I_{E_0}}}\right)\varepsilon_I$\ \, and\ \, $\varepsilon_{I_0}=1.34\times 10^{-7}$.

 \section{Instability of fluid-interfaces in    microscopic or mesoscopic tubes submitted to periodic gravitational forces} 
 
 \subsection{The Mathieu  parametric resonance} Hill's     equation  
 is the  differential equation 
$
 \ddot{x}+f(t)\,x=0  
$,
 where $f(t)$ is a   periodic function.
    Depending of $f(t)$, solutions may stay bounded at all times  or the amplitude of   oscillations  may grow exponentially as described by Floquet's theory \cite{Arnold}.  
 An important  case  is  Mathieu's equation \cite{Mathieu}:   
 \begin{equation}
 \ddot{x}+\omega_0 ^{2}\left(1+\varepsilon \cos (q\,\omega_0\,t)\right)\, x =0 ,
 \label{Matthieu}
 \end{equation}
 where $ \omega_0 \in \mathbb{R}^{+\star}$ is  eigenfrequency of   the system, $ \varepsilon \in \mathbb{R}^{\star} $ is a  small real parameter %
 ($ \left\vert\varepsilon\right\vert \ll 1 $ ) and $q\in \mathbb{R}^{+\star}$  
 (we use the term $\omega_0$ as angular frequency and $\varepsilon$  with the same meaning than in Section III).   This case corresponds to  a differential equation of   motion
 of a pendulum whose frequency $\omega=\omega_0\sqrt{1+\varepsilon \cos (q\,\omega_0\,t)}$ varies over time. The fundamental pendulum period is    $ \displaystyle T_0={2 \pi} / {\omega_0}$. The system of
 Hamilton equations corresponding to  
 \eqref{Matthieu} can be associated with a point of the  plane constituted
 of couples $ \left\{q, \varepsilon \right\}$ (see Fig. \ref{Fig.4}). We have the fundamental Mathieu theorem:\\

 {\bf Theorem:  } 
 \textit{Points of $q$-axis  corresponding to $\varepsilon =0$ 
 are
 stable, except
 points $\displaystyle q= {2}/{p}$, where $\,
 p\in \mathbb{N}$, which are  unstable.}
\\
 
 \noindent The proof of the theorem is given in  \cite{Arnold,Gouin}. 
 As small as $\varepsilon $ is, the theorem proves that the amplitude of oscillations of $x$  exponentially growths when
 $   \displaystyle q= {2}/{p}, \
 p \in \{1, 2\} $.

 The domain of instability of 
 Mathieu's equation in  plane of couples  $\left\{{1}/{q} ,\varepsilon \right\} $ contacts  $ {1}/{q}$-axis at points $ \left\{\displaystyle {1}/{q} =  {p}/{2},\ p\in \mathbb{N}\right\} $. The 
 corresponding values of $  {p}/{2} $ on  $ {1}/{q}$-axis   are called points of
 {\it	parametric resonance}.
 For   plane red domains in Fig.\,\ref{Fig.4}   the solutions of  \eqref{Matthieu} 
 become unstable.
 This 
 parametric resonance is   strongest manifesting when $ p= 1$ or ($q=2$). Parametric resonance   really manifests  for $ q = 2, \, q = 1
 $ and more rarely for $ q = 2/3 $.
  \begin{figure}
 	\begin{center}
 		\includegraphics[width=6cm]{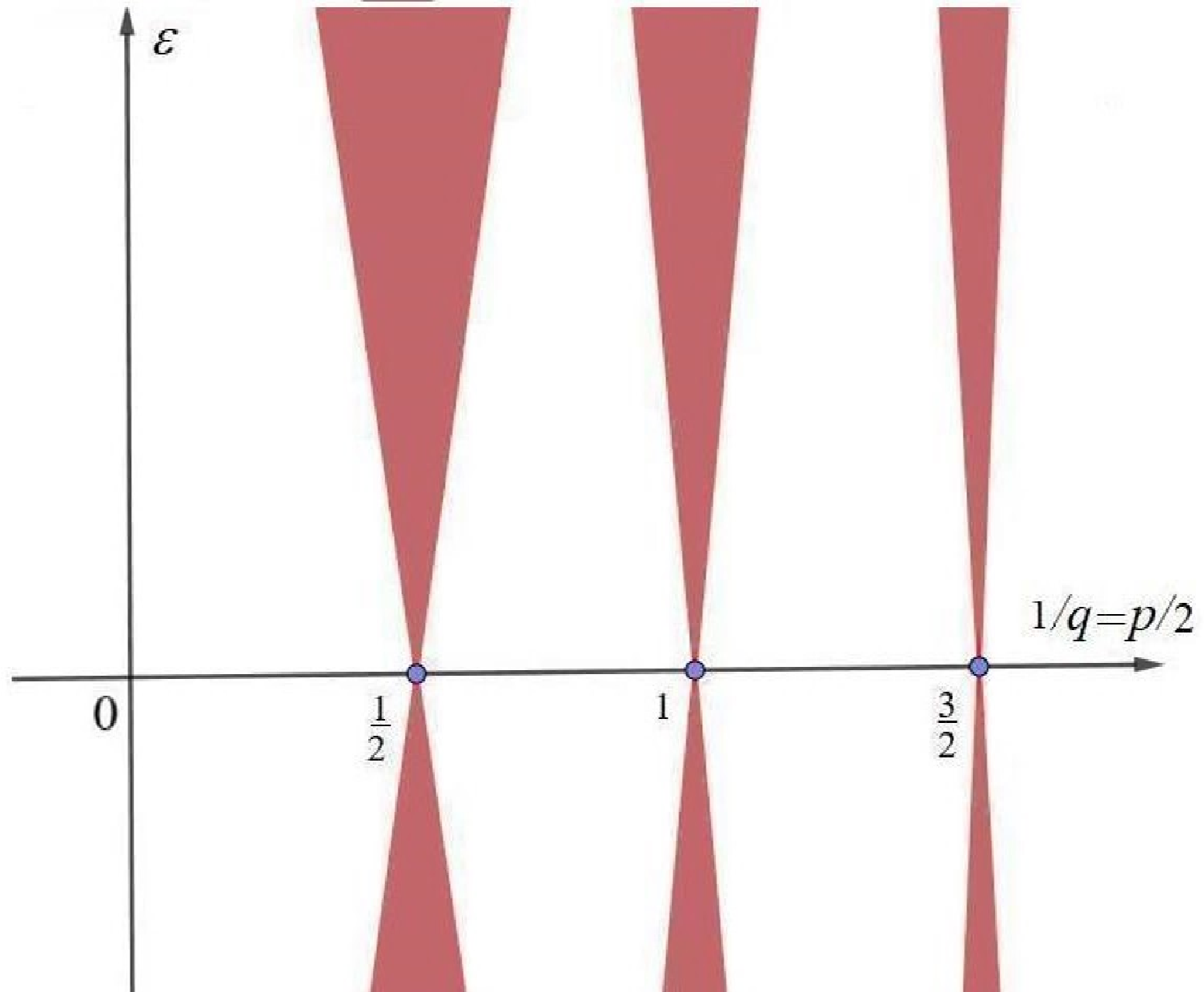}
 	\end{center}
 	\caption{Instability  domains in the parametric resonance case.}  \label{Fig.4}
 \end{figure}

 \subsection{Influence of the lunisolar-tidal's resonance on plants}
 
 As seen in Section 3,      the gravity is not constant during revolutions of the Moon and the ISS around the Earth.
 The liquid considered in     \textit{toy-models} is sap. On Earth,  horizontal  microscopic or mesoscopic
 tubes schematize plant root channels;  vertical tubes schematizes stem channels. Inside the ISS,  \textit{toy-model} of Section II.1  schematizes roots and stems  filled by  sap.   
 The   channels are tubes of xylem or cells     conducting water, crude sap, or elaborated sap, mixed with air-vapor.  Equations \eqref{etapert} and \eqref{ppert} are modified  by    lunisolar-tides to obtain Mathieu's equations.      \\

 \subsubsection{Resonances on the Earth}
 
 From Section 3, \ 
 $\omega_{L_{E_0}}=2\pi\times 1.12\times 10^{-5}$     {\rm rad.s$^{-1}$,
   $\displaystyle T_{L_E}=24.8$ hours 
 and      $\displaystyle \varepsilon_{_{E_0}}   =0.47\times 10^{-8}$.\\
 Perturbations ${\eta}$ associated with    Moon revolutions  around Earth-bound axes    verifies    
$
 \ddot{\eta}+\omega_{L_E}^2 \, \eta =0,
$
 but angular frequency $\omega_{L_E}$ is not constant and the differential equation \eqref{etapert} writes: 
 \begin{equation*}
 \ddot{\eta}+\omega_{L_{_{E0}}}^2  \left(1+\varepsilon_{_{E_0}} {\rm cos} \left(\omega_{L_{E_0}}\,(t-t_0)\right)\right) \, \eta =0\label{Mathieu1}
 \end{equation*}
The perturbations of  pressure through the liquid-sap/air-vapor interface verify the differential equation \eqref{ppert}:
 \begin{equation*}
 [\ddot p]+\omega_{L_{E_0}}^2  \left(1+\varepsilon_{_{E_0}} {\rm cos} \left(\omega_{L_{E_0}}\,(t-t_0)\right)\right) \, [p] =0 \label{Mathieu2}
 \end{equation*}\\
 We are in   the case $q=1$ of Eq. \eqref{Matthieu} corresponding to a small but effective impulsion.

 \subsubsection{Resonances inside the ISS}

 From Section 3, \
 $\omega_{I_{E_0}}=2\pi\times 1.79\times 10^{-4}$ {\rm rad.s$^{-1}$,     $T_{I_E} = 93\ {\rm minutes}$
 and     $\varepsilon_{I_0}=1.34\times 10^{-7}$.\\
 Perturbation ${\eta}$ associated with   revolutions of the ISS around Earth-bound axes  verifies  
$
 \ddot{\eta}+\omega_{I_E}^2 \, \eta =0,
$
 but frequency $\omega_{I_E}$ is not constant and the differential equation \eqref{etapert} writes: 
 \begin{equation*}
 \ddot{\eta}+\omega_{I_{E_0}}^2  \left(1+\varepsilon_{I_0} {\rm cos} \left(2\,\omega_{I_{E_0}}\,(t-t_1)\right)\right) \, \eta =0\label{Mathieu3}
 \end{equation*}
 The perturbations of  pressure through the liquid-sap/air-vapor interface verify the differential equation:
 \begin{equation*}
 [\ddot p]+\omega_{I_{E_0}}^2  \left(1+\varepsilon_{I_0} {\rm cos} \left(2\,\omega_{I_{E_0}}\,(t-t_1)\right)\right)\, [p] =0 \label{Mathieu4}
 \end{equation*}
 We are in   case $q=2$ of Eq. \eqref{Matthieu}  corresponding to a bigger impulsion than for Earth case.  \\
 
The results   come from  periodic gravity variations     of   period commensurable with  period of revolution of the Moon around Earth-bound  axes. 
  It can be seen in  papers by Barlow \textit{et al}  that the effect of gravity  variations   on plants is steeper for the ISS since  bean leaves oscillate dramatically.  This observation is in accordance with Mathieu's equations: the period of revolution of the  ISS is much shorter than     period of revolution of the Moon around the Earth and     coefficient $\varepsilon$ in Mathieu's equation is  35 times larger than the one associated with gravity variations on Earth. In addition, the resonance coefficient $q$  is   optimum   since the period of gravity variations   in the ISS is half  of its circumterrestrial revolution. 
 
 \section{Discussion and conclusion}
 
 The instabilities of fluid interfaces in microscopic or mesoscopic tubes subjected to periodic gravity come from Mathieu's equations.  As the variations of gravity are periodic and commensurable with the periods of revolution of the Moon or the ISS around the Earth's axes, the instabilities always appear during  rotation of the Earth and  revolution of the ISS.   
 \\
 Inside the ISS, it can be seen    that   effects on plants are  stronger than on Earth since   leaves  dramatically oscillate depending along   the ISS' cycle.    
Despite its extreme smallness, the variation of gravity on Earth has great consequences on ocean tides. The effect is less spectacular on plants.   It is   proven that  disturbances of Mathieu's equation exponentially increase  over time \cite{Arnold}; then, the successive rotations of the ISS  and   Moon around the Earth  amplify observed  perturbations.

 The lunar orbit relative to the Earth is not circular but elliptical with apogee of 405 400 km and perigee of 362 600 km. This fact was not taken into account in our calculations because the spacing between the two bodies does not vary significantly during $93$ minutes or one day.  The gravity variation  depending on this distance, we can assume that effects on plants and living organisms depend on the rising Moon and the falling Moon as  \textit{Tavener's  gardene}r thinks \cite{Tavener}. 

 Also at a mesoscopic level, interfacial capillarity is one of the main drivers of the rise of crude  sap in trees \cite{Gouin1,Gouin2}. It   provokes, by pressure variations and time-exponential disturbances,   lengthening of stems and   swelling of roots.   It can be assumed that the same phenomenon occurs in   cells  and thus explains the influence of  lunisolar tide on all living organisms including animals and humans as predicted by \cite{Miles,Gluckman}.
 In astronomy, orbital resonances  appear; the periodic gravitational influence may destabilize the orbits of asteroids \cite{Brouwer}. 
  Thus, in the whole of nature, at   nanoscopic, microscopic, metric, terrestrial and sidereal scales, materials and creatures are subject to gravitational resonance phenomena. 
 
 We have only considered  instabilities due to lunisolar tides without developing information on the dynamics of the induced flows. Such an important development could be the subject of voluminous  numerical computations. 
 One could thus apply the Rayleigh-Taylor model for compressible fluids  \cite{ Bian} or for incompressible fluid turbulence \cite {Boffetta}.
\\

  \textbf {Data Availability Statement}:
 The data that supports the findings of this study are available within the article and its bibliography.\\

 \textbf {Acknowledgments}:
 The author   is partially supported by CNRS, GDR PhyP:  Bio-m\'ecanique et biophysique des plantes  and by Agence Nationale
 de la Recherche, France (project SNIP ANR–19–ASTR–0016–01).\\ I would like to thank Robert Findling,  Sergey Gavrilyuk,  and anonymous referees for their helpful comments and remarks.
  \appendix
\section{Compatibility with equations of motion and boundary conditions}
The purpose of this appendix is to show that the system \eqref{Pert} and the equations \eqref{etapert} and \eqref{ppert} are compatible with the equations of motion \eqref{momentumbis1}.
\subsection{ Interfacial perturbations}
From
 \begin{equation*} 
 dH = \frac{\partial H }{\partial t }dt + {\rm grad}H  \,\boldsymbol{.}\,  d\boldsymbol{M}\quad {\rm with}\quad  d\boldsymbol{M}=\boldsymbol{w}\,dt
 \end{equation*}
 where $\boldsymbol{w}$ is the  fluid velocity   along the interface, we get:
 \begin{equation}
 \frac{\partial H }{\partial t } =-  {\rm grad}H\,\boldsymbol{.}\, \boldsymbol{w} \qquad {\rm with}\qquad \frac{\partial H }{\partial t } =-\frac{\partial \eta }{\partial t } \label{surf}
 \end{equation}
 In cylindrical coordinates:
 \begin{equation*}
 {\rm grad}\,H=\boldsymbol{e}_r-\frac{1}{r}\frac{\partial\eta}{\partial \theta}\boldsymbol{e}_\theta-\frac{\partial\eta}{\partial z}\boldsymbol{e}_z
 \end{equation*}
 where $\boldsymbol{e}_r$, $\boldsymbol{e}_\theta$ and $\boldsymbol{e}_z$ are the unit vectors of    coordinate-lines.   The external unit normal-vector to the interface is: 
 \begin{equation*}
 \boldsymbol{n}=\displaystyle \frac{\displaystyle \boldsymbol{e}_r- \frac{1}{r}\frac{\partial\eta}{\partial \theta}\boldsymbol{e}_\theta-\frac{\partial\eta}{\displaystyle \partial z}\boldsymbol{e}_z}{\displaystyle \sqrt{  1+\frac{1}{r^2}\left(\frac{\partial\eta}{\partial \theta}\right)^2+ \left(\frac{\partial\eta}{\partial z}\right)^2}}
 \end{equation*}
 From  \eqref{surf} we deduce:
 \begin{equation}
 \boldsymbol{w}\,.\,\boldsymbol{n}=\frac{\displaystyle \frac{\partial \eta}{\partial t}}{\displaystyle \sqrt{  1+\frac{1}{r^2}\left(\frac{\partial\eta}{\partial \theta}\right)^2+ \left(\frac{\partial\eta}{\partial z}\right)^2}}\label{normal vel}
 \end{equation}
 For  fluids in $(a)$ and $(b)$ domains, we write:
 \begin{equation*}
 \boldsymbol{u}_j =  \frac{\partial \phi_j}{\partial r}\boldsymbol{e}_r +\frac{1}{r} \frac{\partial\phi_j}{\partial \theta}\boldsymbol{e}_\theta + \frac{\partial\phi_j}{\partial z}\boldsymbol{e}_z \qquad 
 j\in\{a, b\}
 \end{equation*}
 and from  \eqref{surf}, \eqref{normal vel}, we get   interfacial conditions:
 \begin{equation}
 \frac{\partial \phi_j}{\partial r}-\frac{1}{r^2} \frac{\partial\phi_j}{\partial \theta}\frac{\partial\eta}{\partial\theta}
 - \frac{\partial\phi_j}{\partial z} \frac{\partial \eta}{\partial z}=\frac{\partial \eta}{\partial t}\qquad 
 j\in\{a, b\}\label{key}
 \end{equation}
 The sum $\mathcal H$ of principal curvature radii of    liquid-sap/air-vapor interface is ${\mathcal H          }={\rm div} \, \boldsymbol{n}$ \ \cite{Alex}.
 Therefore,
 \begin{eqnarray*}
 	\mathcal H = \frac{2}{\left(r^2+\eta^{\prime 2}_\theta +r^2\eta^{\prime 2}_z\right)^{0.5}} -\frac{r^2\left(1+\eta^{\prime 2}_z\right)}{\left(r^2+\eta^{\prime 2}_\theta +r^2\eta^{\prime 2}_z\right)^{1.5}}\\ -\frac{\eta^{\prime\prime}_{\theta^2}}{r\left(r^2+\eta^{\prime 2}_\theta +r^2\eta^{\prime 2}_z\right)^{0.5}}	+ \frac{\eta^\prime_\theta}{r} \frac{\eta^\prime_\theta\eta^{\prime\prime}_{\theta^2}+r^2\eta^\prime_z\eta^{\prime\prime}_{z\theta}}{\left(r^2+\eta^{\prime 2}_\theta +r^2\eta^{\prime 2}_z\right)^{1.5}} \\ \\ 
 	-\frac{r\eta^{\prime\prime}_{z^2}} {\left(r^2+\eta^{\prime 2}_\theta +r^2\eta^{\prime 2}_z\right)^{0.5}}+ \frac{r\eta^{\prime}_{z} \left(\eta^{\prime }_\theta\eta^{\prime\prime }_{\theta z}+r^2\eta^\prime_z\eta^{\prime\prime}_{z^2}\right)}{\left(r^2+\eta^{\prime 2}_\theta +r^2\eta^{\prime 2}_z\right)^{1.5}}
 \end{eqnarray*}
 By expanding to  first order with respect to $\eta$  expression of $\mathcal H$, we obtain for   total curvature on the perturbed interface:
 \begin{equation*}
 \mathcal H = \frac{1}  {r_a}     -\frac{1}{r_a^2}     \left(\eta+ \eta^{\prime\prime}_{\theta^2}\right)-        \eta^{\prime\prime}_{z^2}   +\mathcal{O}(\eta^2)
 \end{equation*}
 \subsection{Compatibility   with motion equations}
 \subsubsection{The incompressibility}
Equation of incompressibility \eqref{incompresible} yields:
\begin{equation*}
\displaystyle\left(\frac{1}{r}\,\frac{\partial {\hat{\phi_j}} }{\partial r} +\frac{\partial^2{\hat{\phi_j}} }{\partial r^2}+\frac{1}{r^2} \frac{\partial^2 {\hat{\phi_j}} }{\partial \theta^2}-k^2\hat{\phi_j}\right) \, e^{i(kz-\omega t)}=0
\end{equation*}
which is equivalent to:
\begin{equation}
\displaystyle r^2\,\frac{\partial^2 {\hat{\phi_j}} }{\partial r^2} +r\frac{\partial {\hat{\phi_j}} }{\partial r } -k^2r^2\hat{\phi_j}+  \frac{\partial^2  {\hat{\phi_j}} }{\partial \theta^2}  =0\label{perturb}
\end{equation}
Classically, we look for   solutions of  \eqref{perturb} in   form: 
\begin{equation*}
\hat{\phi_j}(r,\theta)= \varphi_j(r)\, \psi_j(\theta)
\end{equation*} 
We obtain:
\begin{equation*}
\frac{  r^2\,{\varphi_j}^{\prime\prime}(r) +r {\varphi_j}^{\prime}(r) -k^2r^2\,{\varphi_j}(r)}{{\varphi_j}(r)}+ \frac{{\psi_j}^{\prime\prime}(\theta) }{{\psi_j}(\theta)}=0
\end{equation*}
$\bullet\quad$ Consequently, there exist  two real constants $\alpha_j, \  j\in\{a, b\}$ such that:
\begin{equation*}
\frac{{\psi_j}^{\prime\prime}(\theta) }{{\psi_j}(\theta)}=-\alpha_j\quad\Longleftrightarrow\quad{\psi_j}^{\prime\prime}(\theta)+\alpha_j{\psi_j}(\theta)=0\label{oscil}
\end{equation*}
Due to the cylindrical geometry of  tubes, ${\psi_j}(\theta)$ must be periodic  with a period which is a divider of $2\pi$. Then, $\alpha_j = n_j^2$ where $n_j\in \mathbb{N}^\star$, and 
\begin{equation*}
{\psi_j}(\theta)= A_j {\rm sin}(n_j(\theta-\theta_{0j}))\quad{\rm with}\quad \theta_{0j}\,\ {\rm and}\,\ A_j\in \mathbb{R}
\end{equation*}
$\bullet\quad$ Furthermore:
\begin{equation}
r^2\,{\varphi_j}^{\prime\prime}(r) +r {\varphi_j}^{\prime}(r) -(n_j^2+k^2r^2)\,{\varphi_j}(r)  = 0\label{key1}
\end{equation}
with $\displaystyle k\,r \leq  k\, r_b$. Due to diameter channel sizes, we assume that $r_b \ll \lambda$, and consequently $k^2r^2\ll1$. So,   \eqref{key1} can be linearized in the form:
\begin{equation}
r^2\,{\varphi_j}^{\prime\prime}(r) +r {\varphi_j}^{\prime}(r) -n_j^2\,{\varphi_j}(r)  = 0\label{key2}
\end{equation}
Solutions of  \eqref{perturb} are:
\begin{equation}
{\hat{\phi_j}}(r,\theta)= \left(A_j\, r^{n_j}+B_j\, r^{-n_j}\right)\, {\rm sin } \left(n_j\left(\theta-\theta_{0j}\right)\right) \label{key5}
\end{equation}
with  $A_j\,\ {\rm and} \,\ B_j\in \mathbb{R}$. Terms 
$
  {\partial \eta}/{\partial \theta},\  {\partial {\phi_j}}/{\partial \theta},\   {\partial \eta}/{\partial z},\  {\partial {\phi_j}}/{\partial z}
$
are small quantities and consequently their products are negligible.
From  \eqref{key} we deduce 
$
 {\partial {\phi_j}}/{\partial r}=  {\partial {\eta}}/{\partial t},\ j\in\{a, b\}
$.
From
\begin{equation*}
\frac{\partial {\phi_j}}{\partial r}= \frac{\partial {\hat{\phi_j}}}{\partial r} \, e^{i(kz-\omega t)}  \quad{\rm and}\quad\displaystyle\frac{\partial {\eta}}{\partial t} = -\eta\, i\,\omega  \equiv -\hat\eta\, i\,\omega\, e^{i(kz-\omega t)}
\end{equation*}
we deduce for $r=r_a$:
\begin{equation}
\  \frac{\partial {\hat{\phi_j}}}{\partial r}  = - \hat\eta \, i\,\omega\qquad\quad  j\in\{a, b\}\label{key3}
\end{equation}
From \eqref{momentumbis1} linearized, 
 $
\rho_j \, ({\partial\phi_j}/{\partial t}) + p_j = 0 
$ and
we obtain:
\begin{equation}
-i\,\omega\,\rho_j{\hat{\phi_j}}+{\hat{p_j}}=0\qquad\quad    j\in\{a, b\}\label{key4}
\end{equation} 
As in \cite{Charru}, to linearize   Laplace's relationship near $r=r_a$,  we consider a Young--Taylor expansion at $\eta =0$. For a  given value of $\theta$, we obtain:
\begin{equation*}
P_j(\eta) = \overline {{P_j}(\eta)}+p_j(\eta) = \overline {{P_j}(0)}+\eta\frac{\partial}{\partial r}\left({{P_j}(\eta)}\right)_{\eta=0}+p_j(0)+\mathcal{O}(\eta^2)
\end{equation*}
From
$
{{P_j}(\eta)}={\overline{{P_j}(0)}}-\rho_jg\, y\equiv {\overline{{P_j}(0)}}-\rho_jg\, r\,{\rm sin}\theta
$,
we get 
$
{\partial{{P_j}(\eta)}}/{\partial r}= -\rho_jg \,{\rm sin}\theta
$
and finally:  
\begin{equation*}
{{P_j}(\eta)}={\overline{{P_j}(0)}}-\eta\,\rho_jg\, {\rm sin}\theta+p_j(0)
\end{equation*} 
We denote $p_a$ and $p_b$ for $p_a(0)$ and $p_b(0)$. Consequently,
\begin{equation*}
P_a(\eta)-P_b(\eta) ={\overline{{P_a}(0)}}-{\overline{{P_b}(0)}}-\eta\,\rho_ag\, {\rm sin}\theta+\eta\,\rho_bg\, {\rm sin}\theta+p_a-p_b
\end{equation*}
From \eqref{Laplace},
$
{\overline{{P_a}(0)}} -{\overline{{P_b}(0)}}=  {\gamma}/{r_a} 
$, and
we obtain the equation:
\begin{equation}
\left(\eta\,\rho_bg \,{\rm sin}\theta-p_b\right)-
\left(\eta\,\rho_ag \,{\rm sin}\theta-p_a\right)=-\gamma\left\{    \frac{1}{r_a^2}     \left(\eta+ \eta^{\prime\prime}_{\theta^2}\right)+        \eta^{\prime\prime}_{z^2}\right\} \label{key6}
\end{equation}
From
$
\left({\partial {\hat{\phi_j}}}/{\partial r}\right)_{r=r_a} =f(r_a)\, {\rm sin}\left(n_j\left(\theta-\theta_ {j0}\right)\right)
$
where $f$ is solution of \eqref{key2} and
  from \eqref{key3}, we obtain:
\begin{equation*}
\hat{\eta}  =h(\omega,r_a)\,{\rm sin}\left(n_j\left(\theta-\theta_ {j0}\right)\right)
\end{equation*}
where $h$ is a convenient function related to $f$. Displacement ${\hat{\eta}}$ being independent of $j \in\{a, b\}$, we have  $n_a=n_b\equiv n$ and $\theta_ {a0}=\theta_ {b0}=\theta_ {0}$. Then,
\begin{equation*}
\hat{\eta }^{\prime\prime}_{\theta^2}=-n^2 h(\omega,r_a)\, {\rm sin}\left(n\left(\theta-\theta_ {0}\right)\right)\quad\Longrightarrow\quad\hat{\eta }^{\prime\prime}_{\theta^2}=-n^2\,\hat{\eta}
\end{equation*}
\subsubsection{Boundary conditions}
In cylindrical coordinates:
\begin{equation*}
\boldsymbol{u}_j = {\rm grad }  \,\phi_j =\frac{\partial \phi_j}{\partial r}\,\boldsymbol{e}_r+\frac{1}{r}\,\frac{\partial \phi_j}{\partial \theta}\,\boldsymbol{e}_\theta+\frac{\partial \phi_j}{\partial z}\,\boldsymbol{e}_z 
\end{equation*}
where $j\in\{a, b\}$, and with expression \eqref{key5}:
\begin{equation*}
\left\{\begin{array}{l}
\displaystyle \frac{\partial{\phi_j}}{\partial r}=    n\,\left(\,A_j\, r^{n-1}-\,B_j\, r^{-n-1}\right)\, {\rm sin } \left(n(\theta-\theta_{0})\right) e^{i(kz-\omega t)}
\\ \\
\displaystyle \frac{1}{r}\,\frac{\partial{\phi_j}}{\partial \theta}=    \left(\,A_j\, r^{n-1}+B_j\, r^{-n-1}\right)\, n\,{\rm cos } \left(n(\theta-\theta_{0})\right) e^{i(kz-\omega t)}
\\ \\
\displaystyle \frac{\partial{\phi_j}}{\partial z}=     \left(\,A_j\, r^{n}+\,B_j\, r^{-n}\right)\, {\rm sin } \left(n(\theta-\theta_{0})\right) i\,k\,e^{i(kz-\omega t)} 
\end{array}\right.
\end{equation*}
The components of $\boldsymbol{u}_a$ must be bounded in $(a)
$ domain. Then $B_a=0$ and
$
\phi_a = A_a\,r^n \, {\rm sin } \left(n(\theta-\theta_{0})\right) e^{i(kz-\omega t)}\ {\rm with}\ n\in \mathbb N^\star
$.
\\
On the    micro-tube's wall  $r =r_b$, the normal component of liquid velocity in domain $(b)$ is zero:
on  $r=r_b$,  $\boldsymbol u_b \, \boldsymbol{.} \,\boldsymbol e_r =0$. 
Then, $\displaystyle \left({\partial {\phi_b}}/{\partial r}\right)_{r=r_b}=0$.\\  From \
$
{\partial{\phi_b}}/{\partial r}=    n\,\left(\,A_b\, r^{n-1}-\,B_b\, r^{-n-1}\right)\, {\rm sin } \left(n(\theta-\theta_{0})\right) e^{i(kz-\omega t)}
$,
we obtain 
$
B_b=A_b\, r_b^{2n}
$. 
Condition \eqref{key3} at $r=r_a$ implies:
\begin{equation*}
\left(\frac{\partial{\phi_a}}{\partial r}\right)_{r=r_a}=\left(\frac{\partial{\phi_b}}{\partial r}\right)_{r=r_a}\,\ \Longrightarrow\,\ A_a\,r_a^{n-1}=A_b\,\left(r_a^{n-1}-r_b^{2n}r_a^{-n-1}\right)
\end{equation*}
and finally with $A=A_b $,  $A_a= A\,\left(1- r_a^{-2n} r_b^{2n}\right)$:  
\begin{equation}
\left\{\begin{array}{l}
\displaystyle  {\phi_a} =    A\,\left(1-r_a^{-2n} r_b^{2n}\right)\,r^n {\rm sin } \left(n(\theta-\theta_{0})\right) e^{i(kz-\omega t)}
\\ \label{instabil}\\
\displaystyle  {\phi_b} =    A\,\left(1+  r^{-2n}r_b^{2n}\right)\,r^n {\rm sin } \left(n(\theta-\theta_{0})\right) e^{i(kz-\omega t)}
\end{array}\right.
\end{equation} 

\subsubsection {Equation of pertubations}

From \eqref{key4} we have
$
\hat{p_a}= i\,\omega\,\rho_a\,\hat{\phi_a}  
$; from \eqref{instabil} we obtain 
$
\hat{\phi_a}= A_a\, r^n {\rm sin } \left(n(\theta-\theta_{0})\right)
$,
and from \eqref{key3}, for $r=r_a$: 
\begin{equation*}
A_a=-\frac{i\,\omega\, \hat{\eta}}{n\,r_a^{n-1}{\rm sin } \left(n\,(\theta-\theta_{0})\right)}
\end{equation*}\\
and from
$
\hat{p_a}= i\,\omega\,\rho_a A_a\, r^n {\rm sin } \left(n(\theta-\theta_{0})\right)
$,
we deduce at $r=r_a$:
\begin{equation*} 
\hat{p_a} =\frac{\omega^2\rho_a}{n}\,r_a\hat{\eta}
\end{equation*}
Likewise,
from \eqref{key4}, we have 
$
\hat{p_b}= i\,\omega\,\rho_b\,\hat{\phi_b}$; 
from \eqref{instabil}:
\begin{equation*}
\hat{\phi_b}= A (r^n+r_b^{2n}r^{-n}) \,{\rm sin } \left(n(\theta-\theta_{0})\right)
\end{equation*}
From \eqref{key3}, for $r=r_a$:
\begin{equation*}
A =-\frac{i\,\omega\, \hat{\eta }}{n(r^{n-1}_a-r_b^{2n}r^{-n-1}_a)\, {\rm sin } \left(n\,(\theta-\theta_{0})\right)}
\end{equation*}\\
From
$
\hat{p_b}= i\,\omega\,\rho_b A\, (r^n+r_b^{2n}r^{-n}) \,{\rm sin } \left(n(\theta-\theta_{0})\right)
$,
we deduce at $r=r_a$:
\begin{equation*} 
\hat{p_b} =\frac{\omega^2\rho_b}{n}
\frac{r_a^n+r_b^{2n}r_a^{-n}}{r_a^{n-1}-r_b^{2n}r_a^{-n-1}}\,
\hat{\eta}
\end{equation*}
and from \eqref{key6}, we obtain the  equation \eqref{mother equation} of II.4 for   possible perturbations  at the interface in cylindrical micro-tubes.

\end{document}